\documentclass[a4paper,11pt]{iopart}

\usepackage{graphicx}
\begin{document} 
\title{Rapidity dependence of saturation in inclusive HERA data with the rcBK equation}


\author{Jan Cepila$^1$ and Jes\'us Guillermo Contreras$^2$}
\address{Faculty of Nuclear Sciences and Physical Engineering, Czech Technical University in Prague, Czech Republic}
\ead{$^1$ jan.cepila@fjfi.cvut.cz, $^2$ jgcn@cern.ch}

\begin{abstract}
The contribution of the different terms of the running-coupling Balistky--Kovchegov (rcBK) equation to the description of inclusive HERA data is discussed. Within this framework an alternative definition of the saturation scale is presented. The definition is based on the ratio of the linear to the non-linear term of the equation. A similar ratio is used to study the contribution of the non-linear term to the evolution of inclusive HERA data with rapidity.  It is found that, although the data are well described, the behaviour of the different terms of the rcBK equation for HERA kinematics is not what it is naively expected from saturation arguments. 
\end{abstract}



\section{Introduction}
\label{sec:intro}

One of the most interesting problems nowadays in  QCD is to understand the high-energy behaviour of the gluon structure of the proton.
Inclusive data from deeply inelastic scattering at HERA show that the proton structure function $F_2(x,Q^2)$ grows rapidly with decreasing values of Bjorken-$x$ at fixed values of the virtuality, $Q^2$, of the photon \cite{Aaron:2009aa}.
At small values of $x$, which correspond to the high-energy limit of QCD, the main contribution to $F_2(x,Q^2)$ comes from the gluon density. The growth of the gluon density for decreasing  $x$ at a fixed scale was predicted in \cite{Fadin:1975cb,Lipatov:1976zz,Kuraev:1976ge,Kuraev:1977fs,Balitsky:1978ic}, where it was understood to be due to gluon-branching processes. 
The unitarity of the cross section implies that the gluon density should stop growing at some point. This is known as saturation. It was predicted to occur in QCD in the seminal work \cite{Gribov:1984tu}, where a nonlinear term was added to the evolution equation of the gluon density. The presence of saturation  generates a hard scale, the so called saturation scale, $Q_s(x)$, which allows to perform perturbative calculations, in spite of the high density of the system. 

A framework to study saturation phenomena is the Color Glass Condansate (CGC); an effective theory, which describes the high energy limit of QCD  \cite{McLerran:1993ni,McLerran:1993ka}.
The CGC generates a set of equations known as the JIMWLK equations \cite{JalilianMarian:1997gr,JalilianMarian:1997dw,Weigert:2000gi,Iancu:2000hn,Iancu:2001ad,Ferreiro:2001qy}.
 These equations are equivalent to a hierarchy of equations found by Balitsky  \cite{Balitsky:1995ub,Balitsky:1998kc,Balitsky:1998ya,Balitsky:2001re}. 
 In the limit of a large number of colours, the hierarchy reduces to one equation, which was independently derived by Kovchegov \cite{Kovchegov:1999yj,Kovchegov:1999ua} within the colour dipole model \cite{Nikolaev:1990ja,Nikolaev:1991et,Mueller:1989st}. 
 This equation, called the Balitsky--Kovchegov (BK) equation, is simpler than the JIMWLK equations, but it already embodies most of the  relevant behaviour as shown by a comparison of numerical solutions in both approaches \cite{Rummukainen:2003ns,Kovchegov:2008mk}. 
Corrections to include the running of the coupling, $\alpha_s$, were derived in \cite{Kovchegov:2006vj, Balitsky:2006wa} and the resulting equation, called rcBK, was solved numerically in \cite{Albacete:2007yr}. The solutions of the rcBK equation have been used to describe a variety of HERA data: inclusive 
\cite{Albacete:2009fh,Albacete:2010sy,Berger:2011ew}, diffractive \cite{Betemps:2009ie} and exclusive production of vector mesons  \cite{Goncalves:2010ci, Berger:2012wx}. Furthermore, the rcBK equation includes naturally geometric scaling, one of the most striking discoveries with HERA data \cite{Stasto:2000er}. Due to these successes the rcBK equation is one of the main tools used today to investigate the phenomenology of saturation in QCD.

Here, we use the rcBK equation to study the rapidity dependence of saturation in inclusive HERA data.
 In section \ref{sec:rcBK} we introduce the rcBK equation, review the numerical method to solve it and describe the behaviour of its solution, the dipole scattering amplitude $N(r,Y)$, as a function of rapidity $Y$ and transverse size $r$ of the dipole. We also discuss an alternative definition of the saturation scale and compare its evolution in rapidity 
  to that of the usual definition. In section \ref{sec:F2} we present $F_2(x,Q^2)$ in the colour dipole formalism and compare it to HERA data. We also analyse the evolution in rapidity of the contribution of the non-linear term to the description of $F_2(x,Q^2)$. In section \ref{sec:GS} we comment on the geometric scaling properties of the dipole scattering amplitude and of the inclusive cross section for the scattering of a virtual photon off a proton. In section \ref{sec:dis} we discuss the implications of our results. Finally, in section \ref{sec:SC} we summarise our findings.

\section{The dipole scattering amplitude in the rcBK framework}
\label{sec:rcBK}

The BK equation describes the evolution in rapidity $Y$ of the scattering amplitude $N(r,Y)$ for the scattering of a colour dipole of transverse size $r$ with a target. During the evolution the dipole may split into two dipoles or two dipoles may recombine into one. Here we follow the work of \cite{Albacete:2007yr,Albacete:2009fh,Albacete:2010sy} and study the equation without impact parameter dependance.

\subsection{The rcBK equation}

The running coupling Balitsky--Kovchegov equation \cite{Kovchegov:2006vj, Balitsky:2006wa} is 

\begin{equation}
\fl\frac{\partial N(r,Y)}{\partial Y}=\int d \vec{r}_1 K(\vec r,\vec r_1,\vec r_2)\Bigg(N(r_1,Y)+N(r_2,Y)-N(r,Y)-N(r_1,Y)N(r_2,Y)\Bigg)
\label{eq:rcBK}
\end{equation}
where $\vec{r}_2 = \vec{r}-\vec{r}_1$; $r=|\vec{r}|$ and similarly for $r_1$ and $r_2$. The kernel incorporating the running of the coupling is given by

\begin{equation}
K(\vec r,\vec r_1,\vec r_2)=\frac{\alpha_s(r^2) N_C}{2\pi}\Bigg(\frac{r^2}{r_1^2 r_2^2}+\frac{1}{r_1^2}\left(\frac{\alpha_s(r_1^2)}{\alpha_s(r_2^2)}-1\right)+\frac{1}{r_2^2}\left(\frac{\alpha_s(r_2^2)}{\alpha_s(r_1^2)}-1\right)\Bigg),
\end{equation}
with

\begin{equation}
\alpha_{s}(r^2)=\frac{4\pi}{(11-\frac{2}{3}N_f)\ln\left(\frac{4C^2}{r^2\Lambda^2_{QCD}}\right)}
\end{equation}
where $N_f$ is the number of active flavours and $C$ is a parameter to be fixed by comparing to data.

The rcBK equation can be separated into the linear,
\begin{equation}
I_L(r,Y) \equiv \int d \vec{r}_1 K(\vec r,\vec r_1,\vec r_2)\Bigg(N(r_1,Y)+N(r_2,Y)-N(r,Y)\Bigg),
\end{equation}
and the non-linear part,
\begin{equation}
I_{NL}(r,Y) \equiv \int d \vec{r}_1 K(\vec r,\vec r_1,\vec r_2)N(r_1,Y)N(r_2,Y).
\end{equation}
When the contribution of $I_{NL}$ is small with respect to that of $I_L$, the evolution in rapidity of the dipole scattering amplitude is driven by the linear term. When both terms are of the same size, and equilibrium is reached, there is no more evolution and the scattering amplitude is said to be saturated.

The final ingredient to the rcBK equation is the initial condition. 
 For the initial form of the dipole scattering amplitude we use the McLerran-Venugopalan model \cite{McLerran:1997fk}: 
 \begin{equation}
N(r,Y=0)=1-\exp\Bigg(-\frac{\left(r^2Q^2_{s0}\right)^\gamma}{4} \ln \left(\frac{1}{r\Lambda_{QCD}}+e\right)\Bigg)
\end{equation}
with the values of the parameters $Q^2_{s0}$, $C$ and $\gamma$ taken from fit ($e$) in Table 1 of \cite{Albacete:2010sy}. Note that in the fit the initial rapidity, $Y= 0$, is at  $x_0 = 0.01$, where the relation between $Y$ and $x$ is $Y=\ln(x_0/x)$. Within this prescription rapidity $Y=7$ corresponds to $x\approx10^{-5}$; that is,  to the smallest $x$  measured at HERA for perturbative scales. The fit yielded the following values for the parameters: $Q^2_{s0} = 0.165$ GeV$^2$, $\gamma=1.135$ and $C = 2.52$. The fit was performed under the assumption that $\alpha_s(r^2)$ freezes for values of $r$ larger than $r_0$ defined by $\alpha_s(r^2_0)=0.7$. We set $\Lambda_{QCD}$ to 241 MeV. 

\subsection{The dipole scattering amplitude}

The rcBK equation can be solved numerically using Runge--Kutta (RK) methods in parallel over a grid in $r$. 
This approach has been followed in \cite{Albacete:2007yr,Albacete:2009fh,Albacete:2010sy} using a second order algorithm. 
We have solved the rcBK equation using  the following RK methods: forward Euler (order one), Heun (order two) and  Classical  (order four). In all cases we have a grid of 800 points in $r$, spaced uniformly in $\ln(r)$ from $r=10^{-6}$  to 100 GeV$^{-1}$. The equation was solved using steps of 0.05 rapidity units. The formulae for the RK methods are explicitly given in \ref{sec:app}.

\begin{figure}[tbp]
\centering 
\includegraphics[width=0.95\textwidth]{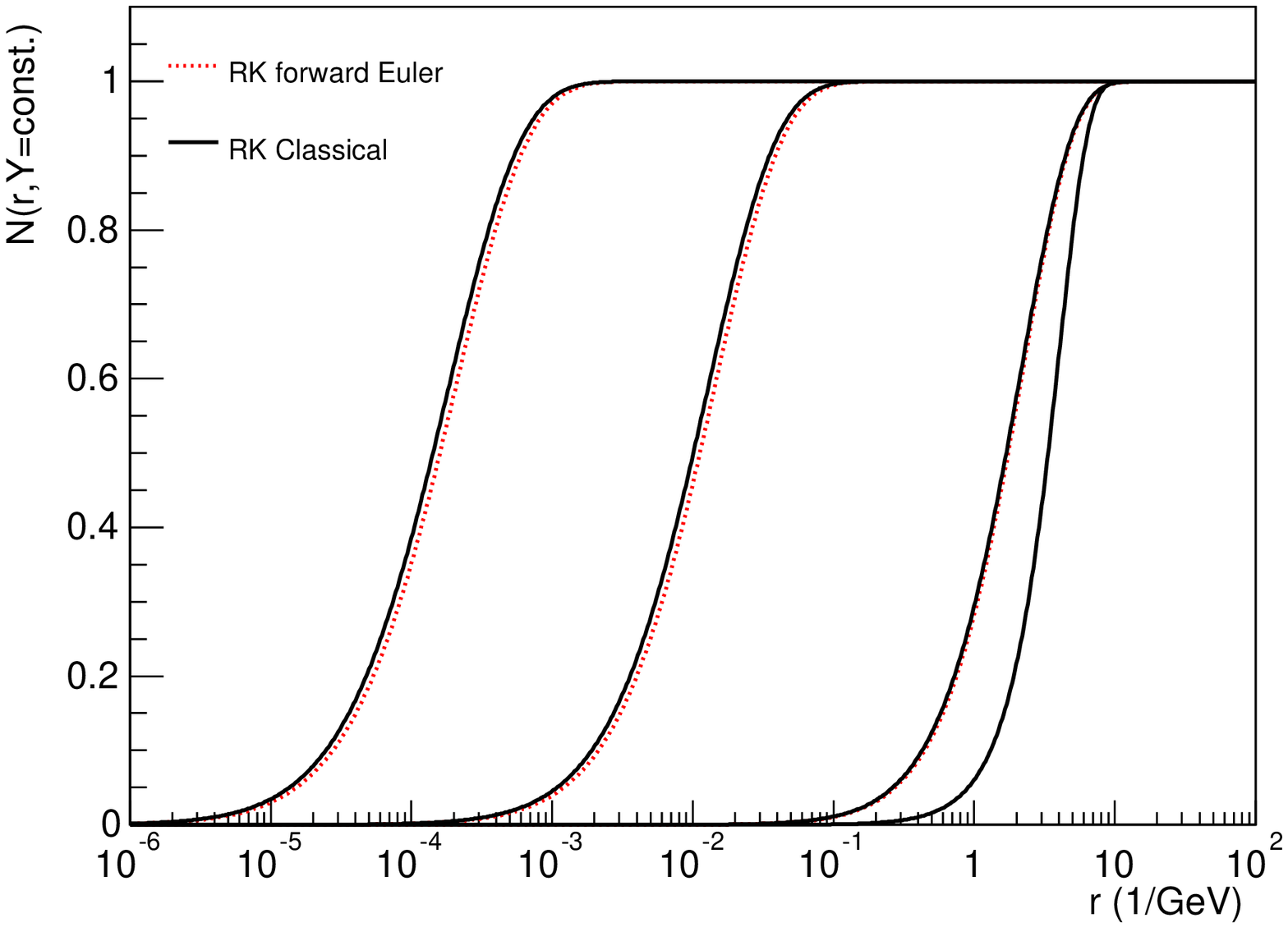}
\includegraphics[width=0.95\textwidth]{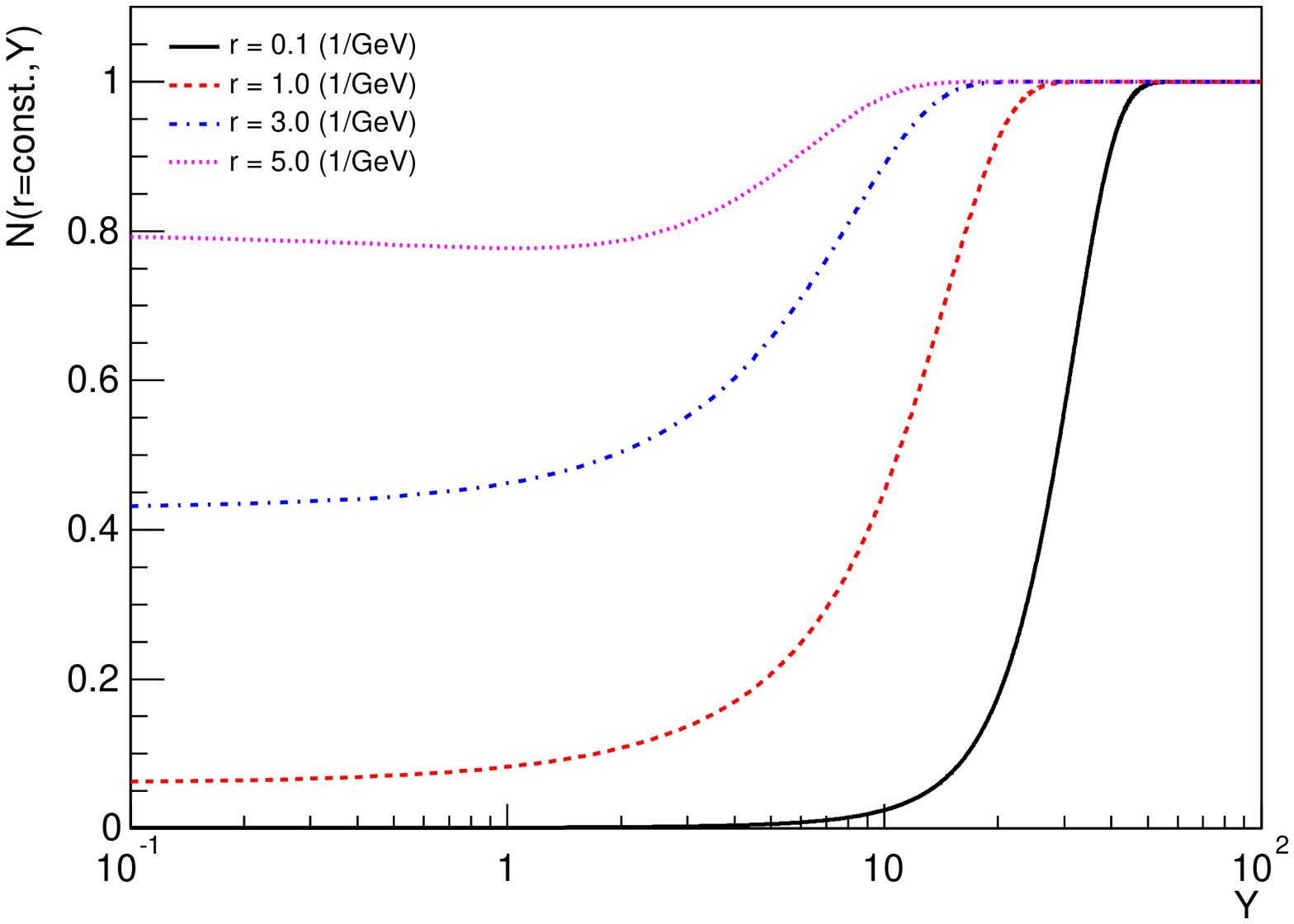}
\caption{\label{fig:N} The solution of the rcBK equation is shown. Upper panel: dipole scattering amplitude $N(r,Y)$ as a function of $r$ at rapidities $Y=0$, 7, 50 and 100 (from right to left)  computed with the  forward Euler and  the Classical methods.  Lower panel: dipole scattering amplitude $N(r,Y)$, found with the Classical method, as a function of rapidity at several values of $r$.}
\end{figure}

The upper panel of Figure \ref{fig:N} shows the solution of the rcBK equation as a function of $r$ at different rapidities for the  RK methods of order one and four. (Heun's method yields results indistinguishable from those of the Classical method, thus it has not been included in the figure.)  
The forward Euler method lags slightly behind the Classical method.  It has been checked that our results agree with those of \cite{Albacete:2010sy}, where a second order RK method has been used. All following  results are based on the Classical method.
The lower panel of Figure \ref{fig:N} shows the dipole scattering amplitude as a function of rapidity at several fixed dipole sizes. The dipole scattering amplitude at a fixed $r$ shows a fairly constant behaviour from $Y=0$ up to a rapidity which depends on the value of $r$, and from there onwards it raises steeply with $Y$ until it saturates. For values of $r$ smaller than around 0.1 GeV$^{-1}$, the growth of $N(r=const.,Y)$ with rapidity starts beyond the kinematic reach of HERA. Even for $r=5$ GeV$^{-1}$, $N(r=5\; {\rm GeV}^{-1},Y)$ is fairly constant down to 
$x\approx 5\cdot 10^{-4}$.

\subsection{The saturation scale}

In order to study the relative importance of the linear to  the non-linear contributions to the evolution of the dipole scattering amplitude with rapidity we define the ratio

\begin{equation}
R_{\rm NL2L} (r,Y) \equiv \frac{I_{NL} (r,Y)}{I_L (r,Y)}.
\label{eq:R2S}
\end{equation}

\begin{figure}[tbp]
\centering 
\includegraphics[width=0.95\textwidth]{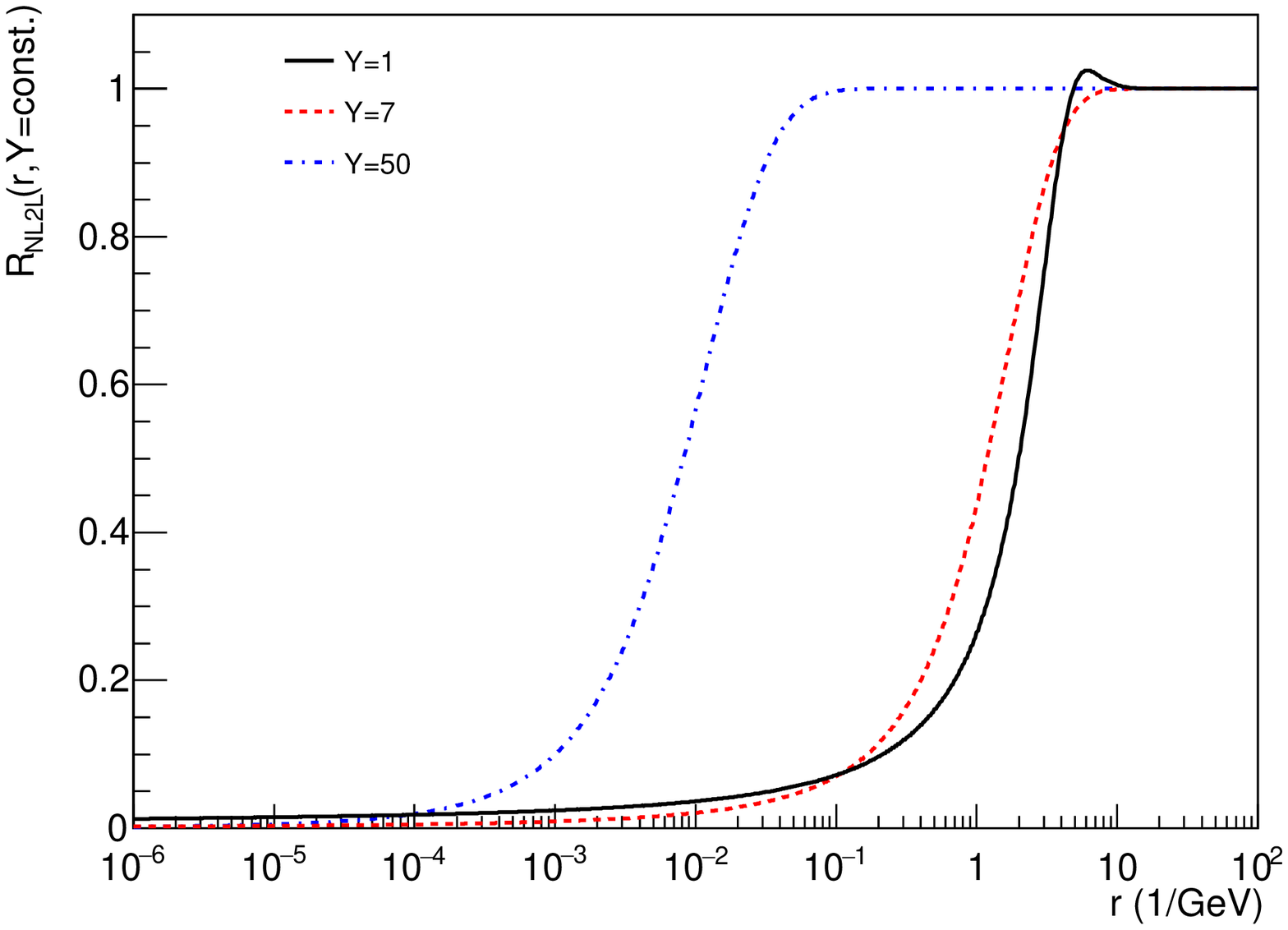}
\includegraphics[width=0.95\textwidth]{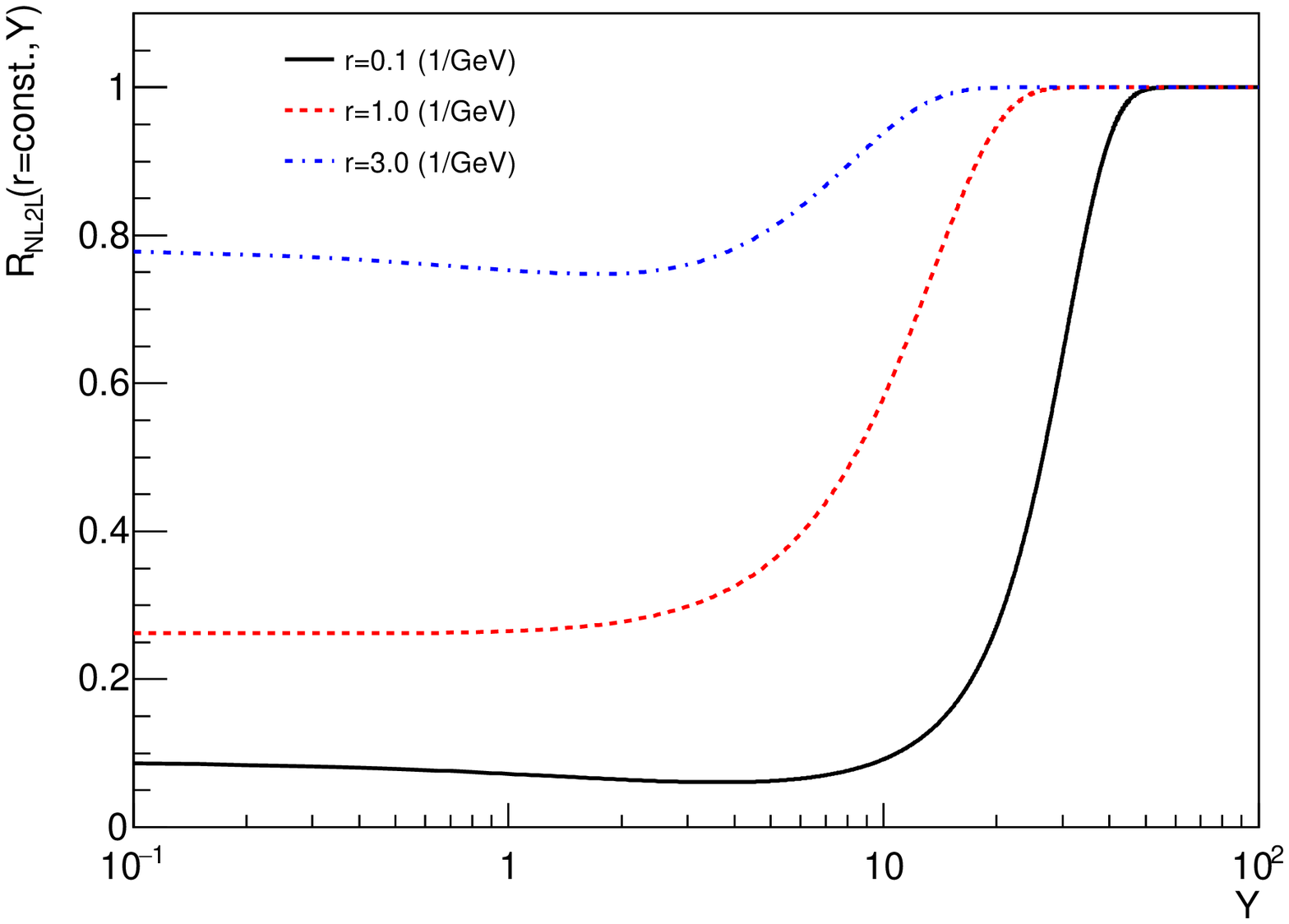}
\caption{\label{fig:R2S} The relative contribution of the non-linear to the linear term of the rcBK equation to the evolution of the dipole scattering amplitude with rapidity is shown as a function of $r$ at constant rapidity (upper panel) and as a function of $Y$ at constant $r$ (lower panel).}
\end{figure}

The upper panel of Figure \ref{fig:R2S} shows $R_{\rm NL2L}$ as a function of $r$ for three different values of rapidity. In all cases, the ratio rises with increasing $r$. At higher rapidities there is a similar behaviour of the ratio and the dipole scattering amplitude: The shape of $R_{\rm NL2L}$ is the same at different rapidities, but it is shifted towards smaller values of $r$, just like in the upper panel of Figure \ref{fig:N}. The lower panel of Figure \ref{fig:R2S} shows $R_{\rm NL2L}$ as a function of rapidity at fixed values of the dipole size $r$. Again, the behaviour is very similar to that of the dipole scattering amplitude. The ratio is flat or even slightly decreasing up to a value of rapidity, which depends on $r$; from there onwards the ratio rises steeply until it reaches unity. Smaller $r$ have a small ratio at the initial rapidity, while large $r$ has a large ratio.

One commonly used definition of the saturation scale is $N (r=1/Q_s,Y) = \kappa$, where $\kappa$ is an arbitrary value close to one.
  Given the behaviour of the ratio $R_{\rm NL2L}$,  there is another possibility to define $Q_s(x)$, which is perhaps more natural in the context of the rcBK equation: the saturation scale is the inverse of the dipole size, where the ratio of the non-linear to the linear contribution to the rapidity evolution of the scattering amplitude is $\kappa$:

\begin{equation}
R_{\rm NL2L} (r=1/Q_s,Y) = \kappa,
\label{eq:QsR2S}
\end{equation}

\begin{figure}[tbp]
\centering 
\includegraphics[width=0.95\textwidth]{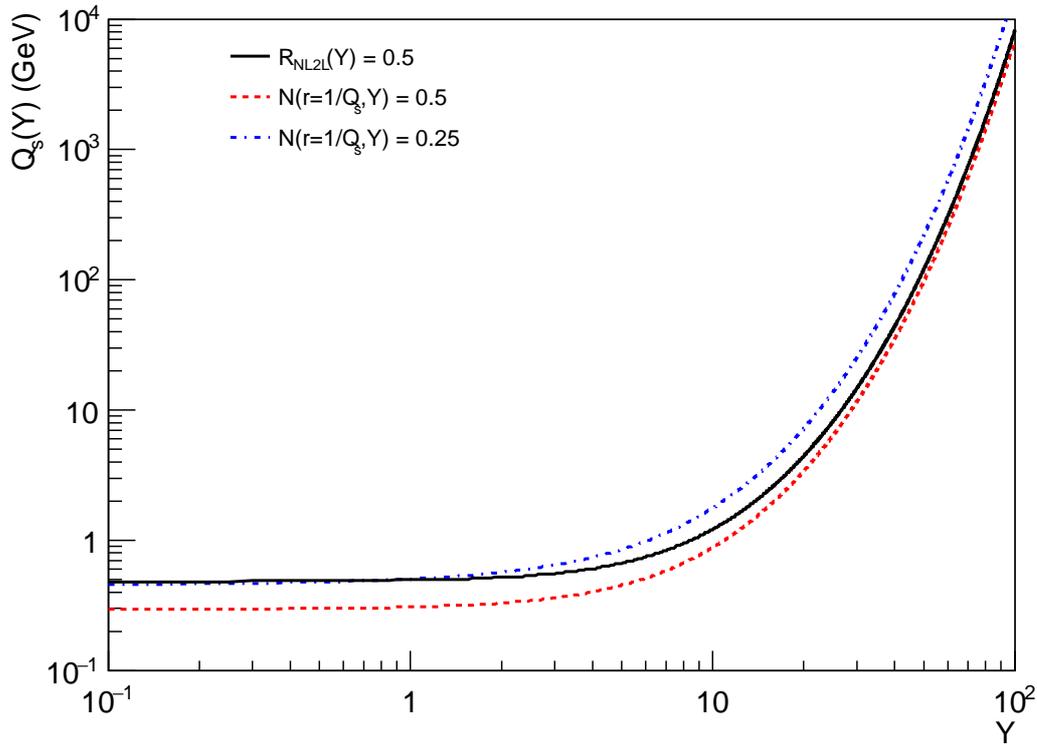}
\caption{\label{fig:Qs} Evolution with rapidity $Y$ of the saturation scale $Q_s(Y)$ for the two definitions discussed in the text.}
\end{figure}

Figure \ref{fig:Qs} compares the evolution with rapidity of the saturation scale for the two definitions. The qualitative behaviour is similar in both cases: the saturation scale  raises very slowly for a few units in rapidity, then there is a large region where $Q_s$  changes faster until it reaches a power-law increase with rapidity, which sets in for $Y \approx $ 20--30.
Quantitatively, at small rapidities, the saturation scale using equation (\ref{eq:R2S}) with $\kappa=0.5$ is very close to the standard definition of the saturation scale  with $\kappa=0.25$, while at large rapidities is closer to the standard definition with $\kappa=0.5$.

It is clear then that both definitions are qualitatively equivalent. The commonly used definition is more general, while the one we introduce here is specific to equation (\ref{eq:rcBK}). As this latter definition is tailored to the rcBK equation, it is helpful to understand qualitatively some of the properties of equation (\ref{eq:rcBK}).

\section{Description of inclusive HERA data in the rcBK framework}
\label{sec:F2}

\subsection{\boldmath The proton structure function $F_2(x,Q^2)$ in the colour dipole formalism}

The proton structure function $F_2$ is related to the cross section $\sigma^{\gamma^\ast p}$ for the scattering of a virtual photon $\gamma^*$ off a proton $p$ by:
\begin{equation}
F_2(x,Q^2)=\frac{Q^2}{4\pi^2\alpha_{em}}\Bigg(\sigma^{\gamma^\ast p}_{T}(x,Q^2)+\sigma^{\gamma^\ast p}_{L}(x,Q^2)\Bigg),
\label{eq:F2}
\end{equation}
where the subindices $T$ and $L$ refer to the transverse, respectively longitudinal, polarisations of  the virtual photon.

In the colour dipole picture \cite{Nikolaev:1990ja,Nikolaev:1991et,Mueller:1989st}, the incoming photon splits into a quark-antiquark pair, which forms a colour dipole, and this dipole interacts with the target. For the case of no impact-parameter dependence the cross section is given   by

\begin{equation}
\sigma^{\gamma^\ast p}_{T,L}(x,Q^2)= \sigma_0 \int d \vec{r} \int^1_0 dz \big |\Psi^{\gamma^\ast\rightarrow q\bar q}_{T,L}(z,r,Q^2)\big |^2 N(r,\tilde{x}).
\label{eq:Sgp}
\end{equation}
To lowest order in $\alpha_{em}$ the square of the wave functions describing the splitting of the virtual photon into the colour dipole are given by

\begin{equation}
\big |\Psi^{\gamma^\ast\rightarrow q\bar q}_{T}(z,r,Q^2)\big |^2=
\frac{3\alpha_{em}}{2\pi^2}\sum_f e_f^2( (z^2+(1-z)^2)\epsilon^2K_1^2(\epsilon r)+m_f^2K^2_0(\epsilon r)),
\end{equation}
and
\begin{equation}
\big |\Psi^{\gamma^\ast\rightarrow q\bar q}_{L}(z,r,Q^2)\big |^2=
\frac{3\alpha_{em}}{2\pi^2}\sum_f e_f^2( 4Q^2z^2(1-z)^2K_0^2(\epsilon r)),
\end{equation}
where we have follow the normalisation convention of \cite{GolecBiernat:1998js}. In these equations, $z$ is the momentum fraction of the photon carried by the quark, $K_{0,1}$ are Macdonald functions, $e_f$ is the electric charge of the quark of flavour $f$ in units of the electron charge and 

\begin{equation}
\epsilon^2=z(1-z)Q^2+m_f^2.
\end{equation}
Only three quark flavours have been considered. For each flavour, the parameter $m_f$ has been fixed at 140 MeV/c$^2$ and the value of $\sigma_0 = 32.895$ mb has been obtained by the fit mentioned above. Finally, following the convention introduced in \cite{GolecBiernat:1998js}, which allows to have a smoother approach to the photoproduction limit, we set

\begin{equation}
\tilde{x} = x\Bigg( 1+\frac{4 m^2_f}{Q^2}\Bigg).
\label{eq:tilde_x}
\end{equation}

\begin{figure}[tbp]
\centering 
\includegraphics[width=0.95\textwidth]{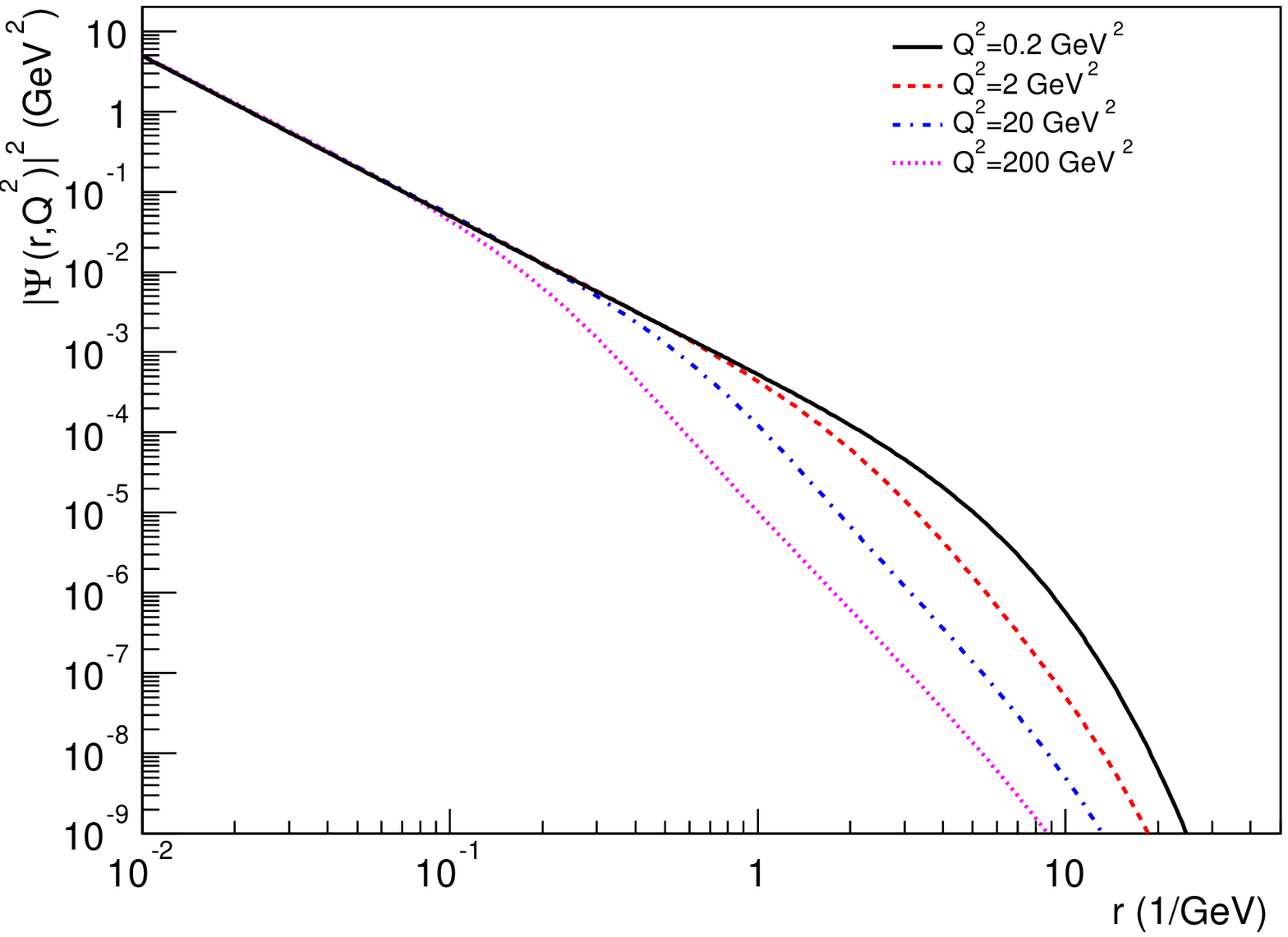}
\includegraphics[width=0.95\textwidth]{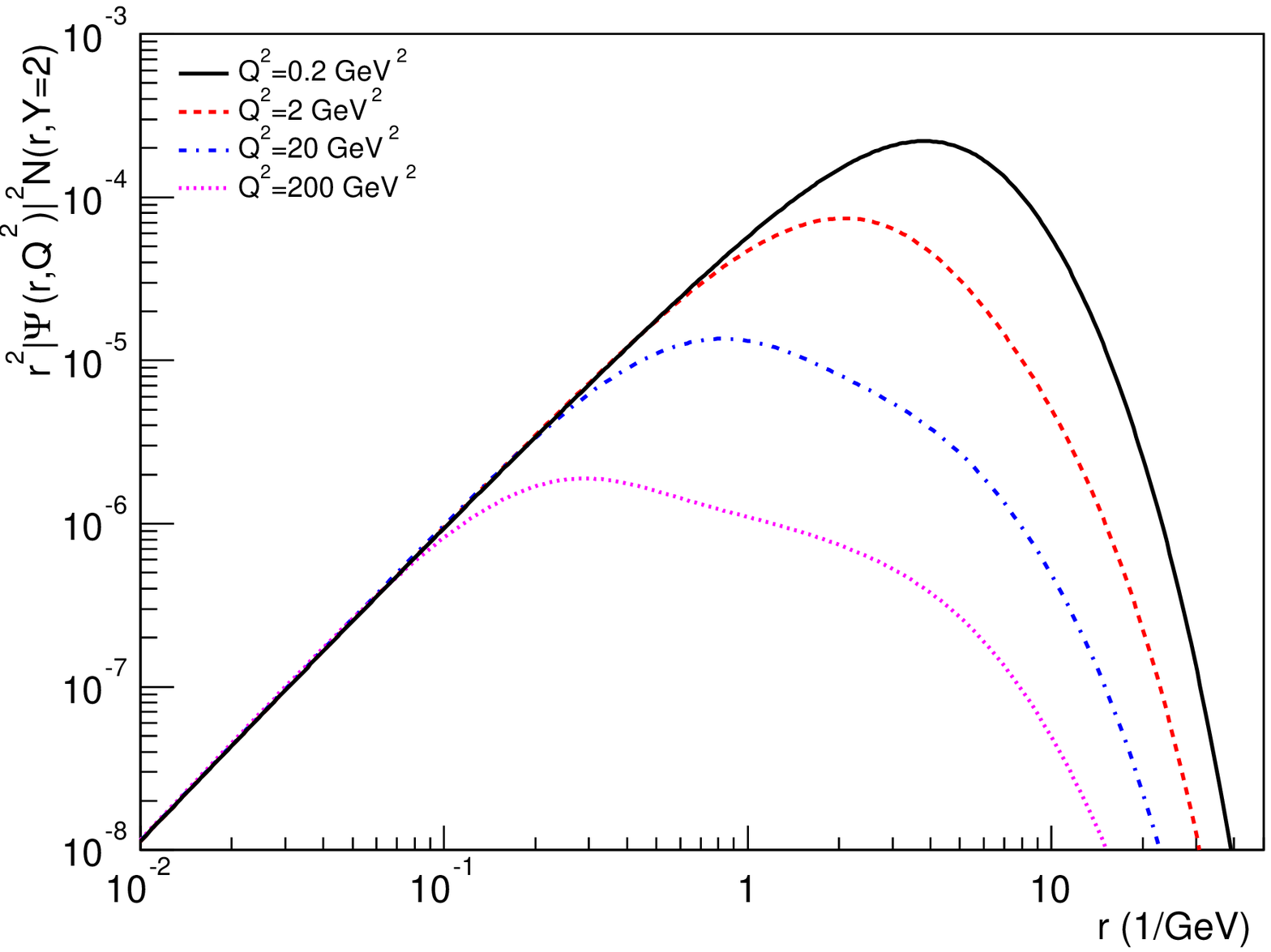}
\caption{\label{fig:WF} Upper panel: The sum of the transversal and longitudinal wave functions integrated over $z$, see equation (\ref{eq:Psi}), is shown for four different values of the virtuality of the photon, $Q^2$.
Lower panel: The integrand of equation (\ref{eq:Sgp}) at $Y=2$ is shown for different values of $Q^2$. }
\end{figure}

The upper panel of Figure \ref{fig:WF} shows
\begin{equation}
\big |\Psi(r, Q^2)\big |^2\equiv \int^1_0 dz\Bigg( \big |\Psi^{\gamma^\ast\rightarrow q\bar q}_{T}(z,r,Q^2)\big |^2 + \big |\Psi^{\gamma^\ast\rightarrow q\bar q}_{L}(z,r,Q^2)\big |^2 \Bigg)
\label{eq:Psi}
\end{equation}
for four different values of the  virtuality of the photon, $Q^2$. The wave function grows rapidly for decreasing dipole sizes reaching the same value for all shown values of $Q^2$ at $r\approx0.1$ GeV$^{-1}$. For large dipoles there are sizeable differences among the wave functions corresponding to the different virtualities. The lower panel of the same figure shows the integrand of equation (\ref{eq:Sgp}) at $Y=2$. The factor $r^2$ originates in the Jacobian to go from $d\vec{r}$ to $dr$ and from $dr$ to $d\ln(r)$. As expected, the integrand goes to zero at small $r$ approximately as $r^2$. The position and height of the maximum of the curves depends on $Q^2$: for the smaller value of $Q^2$ the wave function $|\Psi|^2$ peaks at larger dipole sizes than for larger $Q^2$ values. For these larger values of $Q^2$ the shape of $|\Psi|^2$ is less peaked, more broad. Qualitatively, the same behaviour is found at larger rapidites. 

\subsection{\boldmath Rapidity evolution of the non-linear contribution to  $F_2(x,Q^2)$}

\begin{figure}[tbp]
\centering 
\includegraphics[width=0.95\textwidth]{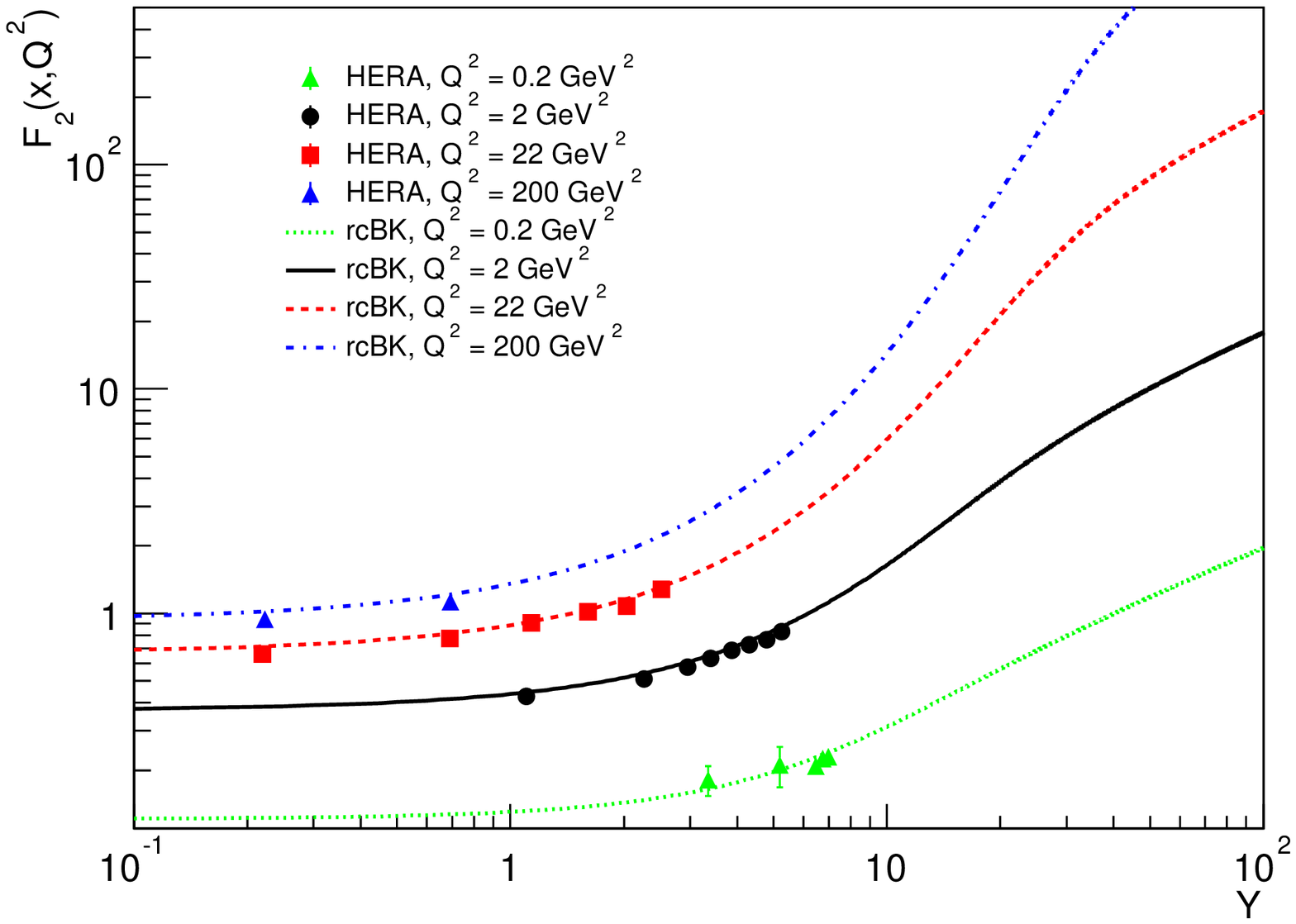}
\includegraphics[width=0.95\textwidth]{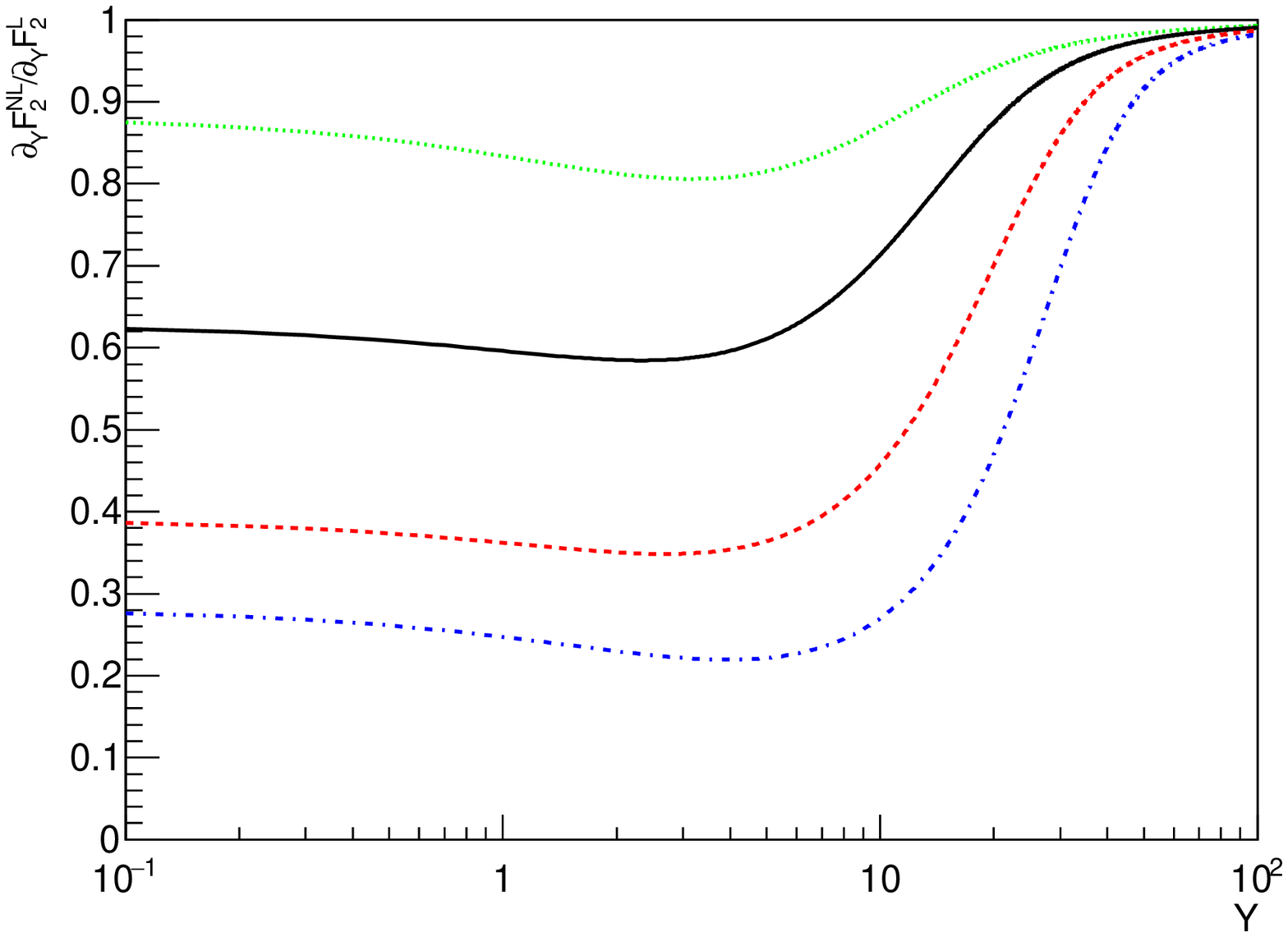}
\caption{\label{fig:F2} Upper panel: description of HERA data from \cite{Aaron:2009aa} using the colour dipole formalism and the solution of the rcBK equation. Here, $Y=\ln(0.01/\tilde{x})$ where $\tilde{x}$ as been defined in equation (\ref{eq:tilde_x}). Lower panel: relative contribution of the non-linear to the linear term of the rcBK equation to the evolution with rapidity of $F_2$ for four different values of the photon virtuality, $Q^2$. Here we have used the notation $\partial_Y \equiv \partial/\partial Y$.}
\end{figure}

The upper panel of Figure \ref{fig:F2} shows the structure function $F_2(x,Q^2)$ as measured by HERA \cite{Aaron:2009aa} for four different values of the photon virtuality $Q^2$ along with its description using the colour dipole formalism and the solution of the rcBK equation. The relation between rapidity and $x$ is given by $Y=\ln(x_0/\tilde{x})$, where $x_0=0.01$ and $\tilde{x}$ as been defined in equation (\ref{eq:tilde_x}). Note that the data at $Q^2 = 200$ GeV$^2$ were not included in the fit, so in a sense the corresponding curve is predicted by the formalism. The description of data by the model is quite good over three orders of magnitude in $Q^2$ and several units of rapidity.
 
Looking at equations (\ref{eq:F2}) and (\ref{eq:Sgp}) it is clear that the evolution of $F_2$ with rapidity is driven by the evolution of the dipole scattering amplitude.
 The left-hand side of equation (\ref{eq:rcBK}) allows to separate the contribution to the evolution of $F_2$ coming from the linear and the non-linear terms. We define $\partial F^{NL}_2/\partial Y$ and $\partial F^L_2/\partial Y$ as the result of exchanging $N(r,Y)$ by $I_{NL}(r,Y)$, respectively $I_L(r,Y)$, in equation (\ref{eq:Sgp}). Their ratio is shown in the lower panel of Figure \ref{fig:F2}. The ratio decreases slightly with rapidity for a few units, and then starts to grow with rapidity towards one. To the lowest values of $Q^2$ corresponds the larger values of the ratio: above 0.5 for $Q^2 =$ 0.2 and 2 GeV$^2$ and below 0.5 for larger values of $Q^2$. 

\section{Geometric scaling}
\label{sec:GS}
\begin{figure}[tbp]
\centering 
\includegraphics[width=0.95\textwidth]{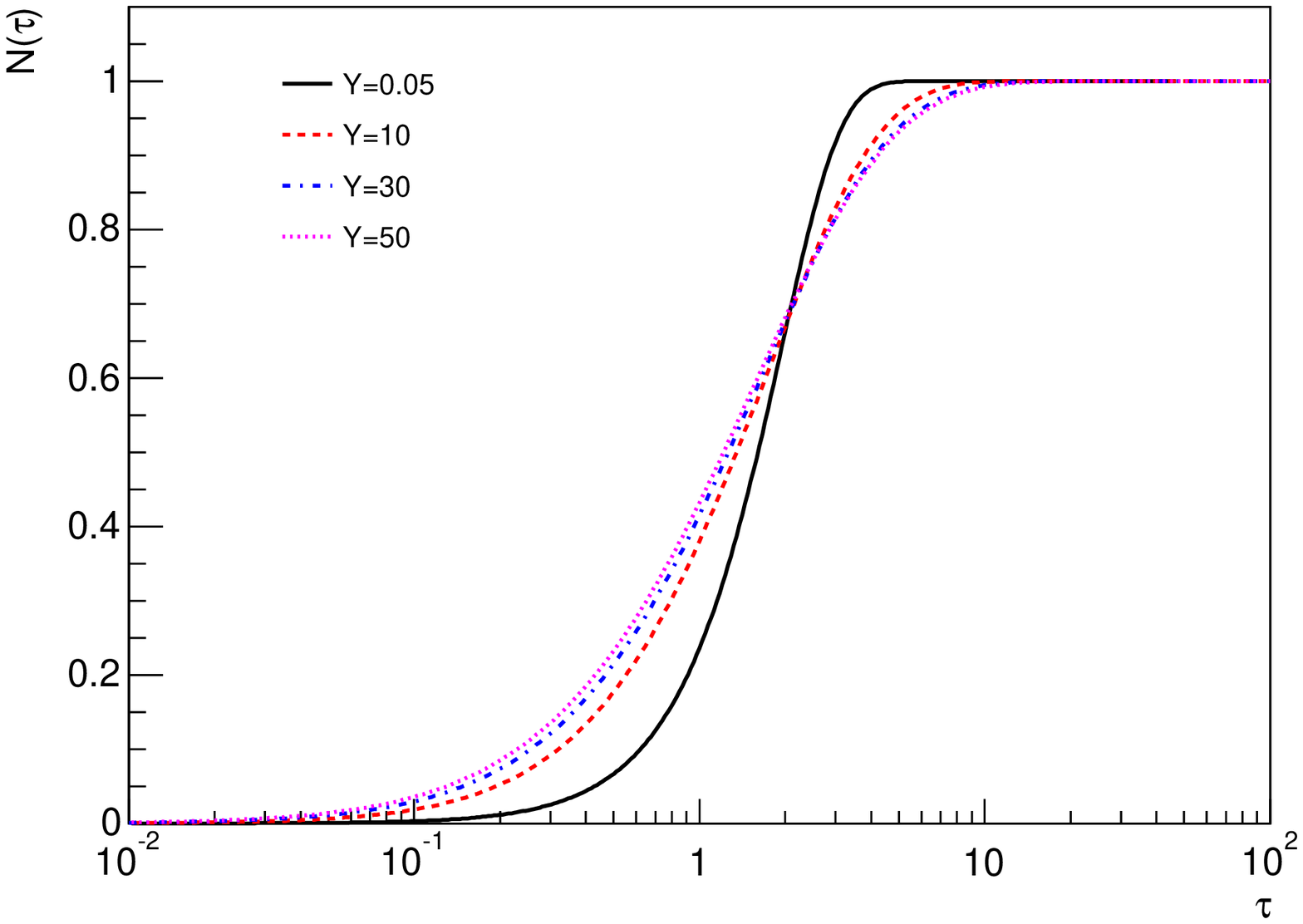}
\includegraphics[width=0.95\textwidth]{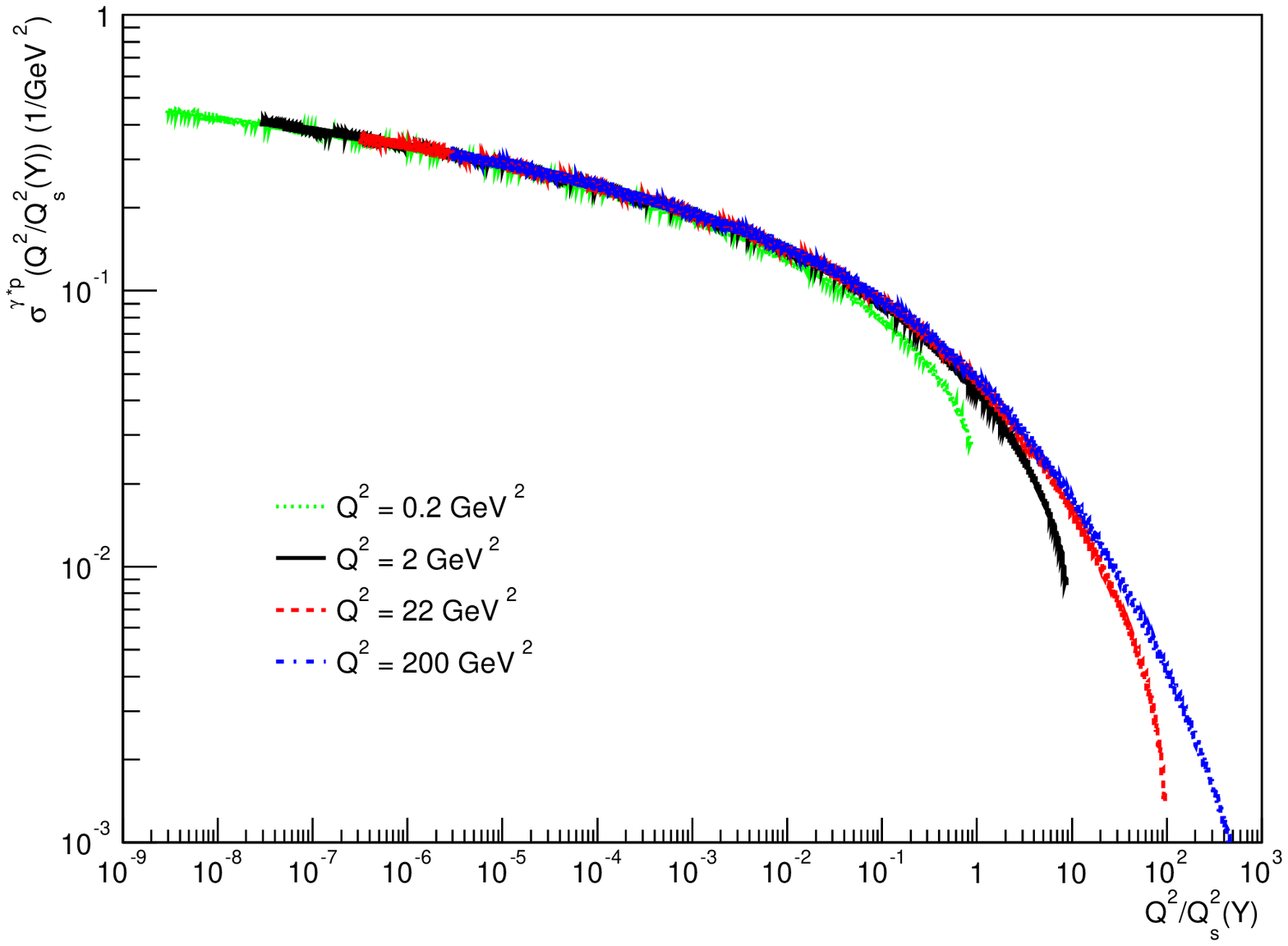}
\caption{\label{fig:GS} Upper panel: The dipole scattering amplitude is shown as a function of $\tau = rQ_s(Y)$ at different rapidities. For the saturation scale, the definition given in equation (\ref{eq:QsR2S}) has been used with $\kappa=0.5$. Lower panel: $\gamma^*p$ cross section as a function of the scaling variable $Q^2/Q^2_s(Y)$ for different values of $Q^2$.}
\end{figure}

One of the most striking phenomena discovered with HERA data is geometric scaling \cite{Stasto:2000er}. On the other hand, the rcBK equation includes naturally geometric scaling in its solutions at large rapidities. It has to be noted that these two geometric scalings are not the same, that is the scaling variables are different, but are intimately related. The upper panel of Figure \ref{fig:GS} shows the dipole scattering amplitude obtained solving the rcBK equation as a function of $\tau = rQ_s(Y)$ at different rapidities. For the saturation scale, the definition given in equation (\ref{eq:QsR2S}) has been used with $\kappa=0.5$. In this case, geometric scaling means 
that the dipole scattering amplitude at different rapidities is the same when plotted as a function of $\tau$. It can be seen that the dipole scattering amplitude at the initial condition is far away from this behaviour. Even at $Y=10$ the dipole scattering amplitude does not yet display geometric scaling. Only for larger values of rapidity the geometric scaling regime is approached. The lower panel of the same figure shows the behaviour of the $\gamma^*p$ cross section as a function of the scaling variable $Q^2/Q^2_s(Y)$. For $Q^2=0.2$ GeV$^2$, $Q^2/Q^2_s(Y)$ is approximately 0.3 at $Y=7$ and grows by a factor of ten by each order of magnitude increase in $Q^2$. The different curves collapse into each other only for values of $Q^2/Q^2_s(Y)$ less than 0.01, that is for rapidities larger than 7.

\section{Discussion}
\label{sec:dis}

The dipole scattering amplitude found with the rcBK equation gives a nice description of HERA data, which has been taken as supporting evidence of saturation effects already for HERA kinematic region. This argument is very strong. Geometric scaling is a striking behaviour, which is naturally found in saturation models. On the other hand the behaviour of a solution of a differential equation has two components, the dynamics inherent to the equation and the influence of the initial condition. 

The rise of the dipole scattering amplitude with rapidity for a fixed dipole size and the appearance of geometric scaling are behaviours associated with the saturation regime. These behaviours appear only at rapidities beyond those reached at HERA when using, in the initial condition, the parameters which allow a good description of data. This behaviour is not found only with this set of parameters. For example, \cite{Albacete:2007yr} reports similar observations when using very different set of parameters for the initial condition.
The saturation scale behaves similarly. The behaviour expected in the saturation regime is reached only at large rapidites.

Turning to the description of data, it is true that the fit is good. But there are two points to consider. First, the fit used data whose rapidity was already very close to the initial condition. This could have given an extra weight to the importance of the initial conditions with respect to the dynamical component of the rcBK equation. Second, data with very low values of $Q^2$ have been included in the fit. (The fit included all available data with $0.045\le Q^2 \le 50$ GeV$^2$.) It is not clear if a perturbative approach is valid at these low scales. It is also not clear how important are these data to determine the values found by the fit. It can be seen in the lower panel of Figure \ref{fig:F2} that it is only for these data at low $Q^2$ that the non-linear term is numerically dominant for the evolution with rapidity, while at larger values of $Q^2$ the evolution is driven mainly by the linear process. Another observation is that the linear process gains in importance with rapidity over the range where data exist,  which is the opposite behaviour of what is naively expected of the dynamics of the rcBK equation.

\section{Summary and conclusions}
\label{sec:SC}

We have studied the contribution of the different terms in the rcBK equation to the description of inclusive HERA data. We find that although the data is well described, the behaviour of the model in the kinematic range of the measurements is not what is expected from the dynamics of the rcBK equation. 
One could try new fits to HERA data without inclusion of the measurements at the lowest $Q^2$ values, which may not be adequate for a perturbative model, and give some time for the evolution of the solution before the first data to be fitted in order to reduce the weight of the initial conditions and to increase the importance of the dynamic content of the rcBK equation.

These results point also to the need of having data at larger rapidities and at perturbative scales. In this context, the measurement of exclusive photoproduction of J$/\psi$ at LHC both off protons \cite{Aaij:2013jxj,Aaij:2014iea,TheALICE:2014dwa} as off nuclei \cite{Abelev:2012ba,Abbas:2013oua,CMS:2014ies} at higher energies than those reached at HERA is very good news. Ultimately the realisation of dedicated electron-hadron colliders \cite{Deshpande:2005wd,Accardi:2012qut,AbelleiraFernandez:2012cc} would provide high quality data to continue these studies and conclude if a saturation regime has been reached.

\ack

This work was partially supported by grant LK11209 of  M\v{S}MT \v{C}R,
 by  grant 13-20841S of the Czech Science Foundation (GACR),
and  by the European social fund within the framework of realising the project ãSupport of inter-sectoral mobility and quality enhancement of research teams at Czech Technical University in PragueÒ, CZ.1.07/2.3.00/30.0034.

\appendix
\section{Runge--Kutta methods to solve the rcBK equation}
\label{sec:app}

Using the following definitions
\begin{equation}
\fl I_0 \equiv \int d \vec{r}_1 K(\vec r,\vec r_1,\vec r_2)\Bigg(N(r_1,Y)+N(r_2,Y)-N(r,Y)-N(r_1,Y)N(r_2,Y)\Bigg),
\end{equation}
\begin{equation}
\fl I_1 \equiv \int d \vec{r}_1 K(\vec r,\vec r_1,\vec r_2)\Bigg(1-N(r_1,Y)-N(r_2,Y)\Bigg),
\end{equation}
\begin{equation}
\fl I_2 \equiv \int d \vec{r}_1 K(\vec r,\vec r_1,\vec r_2),
\end{equation}
the change of $N(r,Y)$ after a step in rapidity $\Delta Y$ can be found using Runge--Kutta methods over a grid in $r$. 

For the forward Euler method, the solution reads:
\begin{equation}
N(r,Y+\Delta Y) = N(r,Y) + \Delta Y I_0.
\end{equation}
For Heun's method, the solution reads:
\begin{equation}
N(r,Y+\Delta Y) = N(r,Y) + \Delta Y I_0 + \frac{(\Delta Y)^2}{2}I_0 I_1 - \frac{(\Delta Y)^3}{2}I^2_0 I_2.
\end{equation}
For the Classical method, the solution reads:
\begin{equation}
N(r,Y+\Delta Y) = N(r,Y) +  \frac{\Delta Y}{6}\Bigg( I_0 +2K_2+2K_3+K_4\Bigg),
\end{equation}
where 
\begin{equation}
K_2 \equiv I_0 + \frac{\Delta Y}{2}I_0 I_1 - \frac{(\Delta Y)^2}{4}I^2_0 I_2,
\end{equation}
\begin{equation}
K_3 \equiv I_0 + \frac{\Delta Y}{2}K_2 I_1 - \frac{(\Delta Y)^2}{4}K^2_2 I_2,
\end{equation}
and
\begin{equation}
K_4 \equiv I_0 + \Delta Y K_3 I_1 + (\Delta Y)^2K^2_3 I_2.
\end{equation}
In our implementation, the values of  the dipole scattering amplitude for sizes $r$, which are not in the grid, are found by a linear interpolation  in $\ln(r)$ between adjacent values of $r$ in the grid. For dipole sizes larger than the larger size in the grid the dipole scattering amplitude has been taken as one, and for dipole sizes smaller than the smaller size in the grid the dipole scattering amplitude has been linearly scaled in $r$ from the value corresponding to the smallest size in the grid.

\section*{References}
\bibliography{rcBKbiblio}

\end{document}